\documentclass{sn-jnl}

\usepackage{graphicx}%
\usepackage{multirow}%
\usepackage{amsmath,amssymb,amsfonts}%
\usepackage{amsthm}%
\usepackage{mathrsfs}%
\usepackage[title]{appendix}%
\usepackage{xcolor}%
\usepackage{textcomp}%
\usepackage{manyfoot}%
\usepackage{booktabs}%
\usepackage{algorithm}%
\usepackage{algorithmicx}%
\usepackage{algpseudocode}%
\usepackage{listings}%

\begin{document}

\title{Multichannel, ultra-wideband Rydberg electrometry with an optical frequency comb}


\author[1]{\fnm{Nikunjkumar} \sur{Prajapati}}\email{nikunjkumar.prajapati@nist.gov}
\author[2]{\fnm{David A.} \sur{Long}}\email{david.long@nist.gov}
\author[1]{\fnm{Alexandra B.} \sur{Artusio-Glimpse}}\email{alexandra.artusio-glimpse@nist.gov}
\author[2,3]{\fnm{Sean M.} \sur{Bresler}}\email{sean.bresler@nist.gov}
\author[1]{\fnm{Christopher L.} \sur{Holloway}}\email{christopher.holloway@nist.gov}

\affil[1]{\orgdiv{Communications Technology Laboratory}, \orgname{National Institute of Standards and Technology}, \orgaddress{\street{325 Broadway}, \city{Boulder}, \state{CO}, \country{USA}}}

\affil[2]{\orgdiv{Physical Measurement Laboratory}, \orgname{National Institute of Standards and Technology}, \orgaddress{\street{100 Bureau Dr.}, \city{Gaithersburg}, \state{MD}, \country{USA}}}

\affil[3]{\orgdiv{Department of Physics}, \orgname{University of Maryland}, \orgaddress{ \city{College Park}, \state{MD}, \country{USA}}}


\abstract{While Rydberg atoms have shown tremendous potential to serve as accurate and sensitive detectors of microwaves and millimeter waves, their response is generally limited to a single narrow frequency band around a chosen microwave transition. As a result, their potential to serve as agile and wideband electromagnetic receivers has not been fully realized. Here we demonstrate the use of a mid-infrared, frequency agile optical frequency comb as the coupling laser for three-photon Rydberg atom electrometry. This approach allows us to simultaneously prepare as many as seven individual Rydberg states, allowing for multichannel detection across a frequency range from 1 GHz to 40 GHz. The generality and flexibility of this method for wideband multiplexing is anticipated to have transformative effects in the field of Rydberg electrometry, paving the way for advanced information coding and arbitrary signal detection.}





\maketitle

\section{Introduction}

Rydberg atoms have become ubiquitous in quantum-based technologies, including entangled ions for quantum computing, efficient microwave to optical conversion, and electric field sensing. Rydberg atoms offer large tuning bandwidths that range from DC to terahertz~\cite{Noah2024,9748947, terehertz_imaging,6910267,waveguide_SA}, are highly sensitive for electric field detection~\cite{Sedlacek2012,Jing2020,doi:10.1063/1.5028357,9069423,Repump_paper,6wavemix,8878963,10.1117/12.2586718,Kumar2017} as well as allowing for direct conversion~\cite{doi:10.1063/1.5099036,doi:10.1116/5.0098057,Wang:17} and phase sensing~\cite{Jing2020, doi:10.1063/1.5028357,8778739,doi:10.1063/1.5088821}. However, unlike classical receivers that continuously monitor the entire spectrum of operation, the instantaneous bandwidth of Rydberg atom sensors are generally limited to tens of MHz around a selected carrier tone~\cite{doi:10.1063/1.5028357,doi:10.1116/5.0098057, Li:22,Song:19,Cox_2018,beam_size_shaffer}. 

Typical sensitive field measurements utilize the resonant Autler-Townes (AT) effect to observe the response of the Rydberg state to external fields~\cite{6910267,first_ryd_elet_shaff}. In this case, the tuning bandwidth is limited to within 50 MHz of the atomic transition and requires slow coarse laser tuning by several nanometers in order to cover the full radiofrequency (RF) reception range~\cite{gallagher_book}. This can be extended to a few hundred megahertz by utilizing complicated Rydberg engineering techniques, but still does not bridge the gap for continuous measurements or multichannel detection of several incoming RF signals~\cite{Simons2021continuous,PhysRevApplied.19.044049}. 

An alternative to the AT effect, the off-resonant Stark shift offers a near continuous tuning capability through the RF spectrum~\cite{waveguide_SA,PhysRevA.107.052605,10.1063/5.0179496}, however, with dramatically reduced typical sensitivity. Additionally, the coupling laser transfers information from multiple RF fields onto a single probe laser in the excitation chain, thus in many cases leading to inseparable signals and limiting the total bandwidth of the system. 
For example, digital television antennas receive RF waves ranging from 50 MHz to 700 MHz and carry information with 6 MHz of modulation bandwidth. By receiving a signal and demodulating the carrier after reception, several signals can be received and readout at the same time. However, with the off-resonant Stark shifting approach, these signals would be inseparable.

For Rydberg atom receivers to compete with classical receivers and realize their full potential, a method for simultaneous measurements of different Rydberg states is critically needed. Here we demonstrate the use of an optical frequency comb as a coupling laser in order to simultaneously prepare seven different Rydberg states, paving the way for multichannel electrometry and sensing across wide frequency bands.

Optical frequency combs have been transformative in wide ranging fields such as time transfer, astronomy, and remote sensing ~\cite{Fortier_Baumann_2019, lehmannreview}. These advancements have largely leveraged the exquisite frequency accuracy that combs can provide as well as their wide spectral bandwidth. However, their low power per comb tooth (generally microwatts or even nanowatts) has proven limiting in some applications ~\cite{Long2023Nanosecond}. Recently we have demonstrated a new approach for spectral translation of electro-optic frequency combs using an optical parametric oscillator (OPO)~\cite{Long2023Nanosecond, heinigerOPO}. Critically, this method allows for highly coherent spectral translation from the near-infrared to the important mid-infrared while also offering exceptionally high power per comb tooth (up to 300 mW). 

Here we utilize this frequency-agile optical frequency comb as the coupling laser in a three-photon Rydberg electrometry instrument, allowing for the preparation of as many as seven Rydberg states simultaneously. Further, we demonstrate that the seven individual states can serve as orthogonal RF channels, allowing for simultaneous and highly sensitive detection of a multitude of RF tones across a range of 1 GHz to 40 GHz, limited only by our available RF sources and horn antennas.

\section{Multi-state Readout}
\begin{figure}[ht]
    \centering
    \includegraphics[width=\columnwidth]{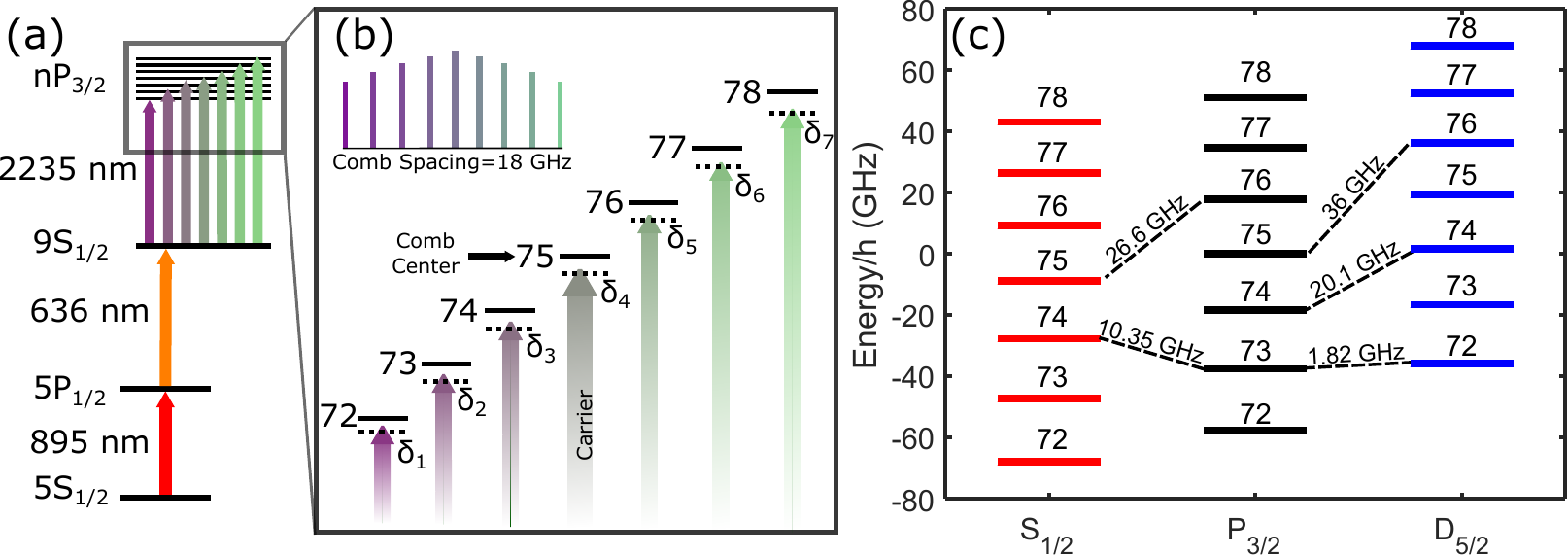}
    \caption{(a) Three-photon excitation to the 75 P$_{3/2}$ state in a Rydberg manifold utilizing 895 nm (probe), 636 nm (dressing), and 2235 nm (coupling) optical fields. (b) Expanded view of the Rydberg manifold which shows the coupling laser optical frequency comb which connects the seven different Rydberg states which are centered around 75 P$_{3/2}$. (c) Energy layout of nearby Rydberg states and some possible RF transition frequencies between the nP$_{3/2}$ states and nearby S and D Rydberg states. }
    \label{fig:LevelAndSample}
\end{figure}

The multi-state Rydberg receiver described here relies upon three-photon excitation to generate and probe Rydberg states in cesium (Cs) atoms~\cite{10.1063/5.0147827,fluoresence} as shown in Fig.~\ref{fig:LevelAndSample}. The three-photon configuration leads to reduced Doppler broadening due to the wavelengths of the various optical fields and therefore a higher theoretical maximum sensitivity~\cite{10.1063/5.0147827}. Although we note that the methods described herein are also applicable to more traditional two-photon approaches~\cite{r13,r12,PhysRevA.102.062817,dual_species}. 

The coherent multi-photon process known as electromagnetically-induced absorption (EIA) allows for probing highly excited Rydberg states with higher precision than other methods such as ionization~\cite{ionization_tilman,gallagher_book,ionization_beam}, electron beam ionization~\cite{ionization_beam}, and direct absorption~\cite{Wang:17}. In the present three-photon process, an 895 nm probe laser, a 636 nm dressing laser, and a 2235 nm coupling optical frequency comb are used in a successive excitation up to the Rydberg state(s) of interest, thus generating a coherent ensemble of atoms. The EIA is a result of this coherence and readout of the probe laser is a measure of the coherence between the ground state and the first excited state. The dressing and coupling laser modify this coherence through their interactions with the 9S intermediate state and the Rydberg state. This projection allows for measurements of Rydberg state(s) of interest by measuring the probe laser transmission. Similarly, the modification of the Rydberg state by an RF field also affects the probe laser transmission.

\begin{figure}[ht]
    \centering
    \includegraphics[width=\columnwidth]{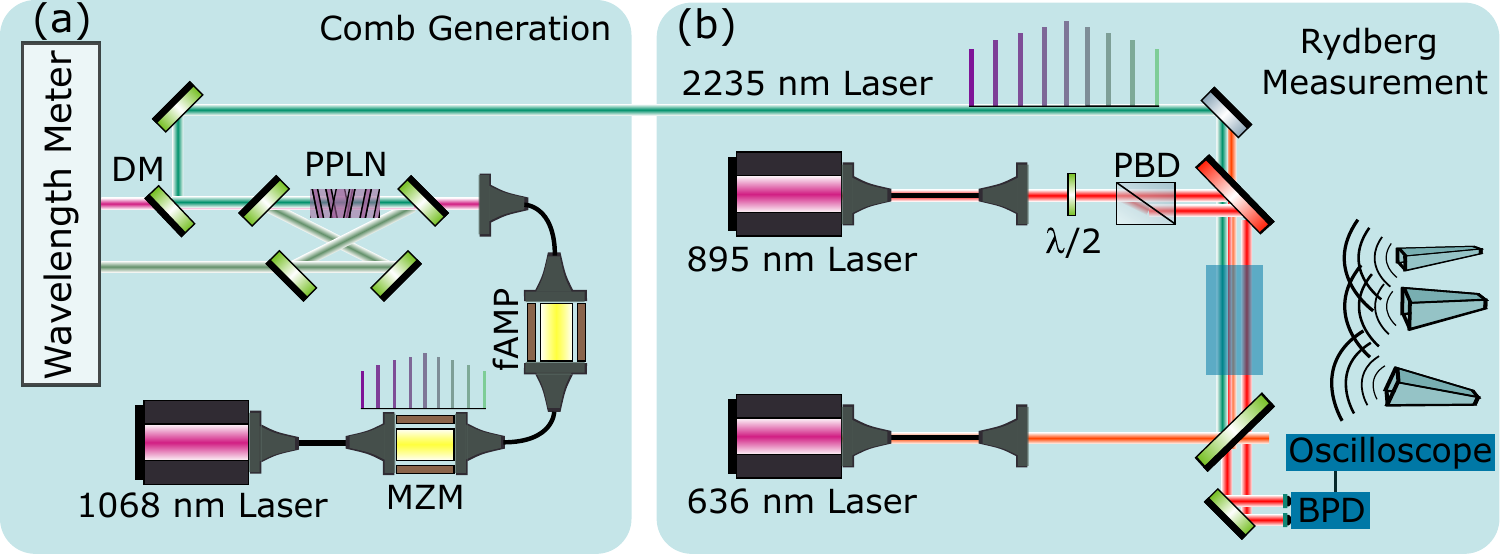}
    \caption{(a) Mid-infrared comb generation schematic and (b) measurement system schematic. The shown acronyms are dichroic mirror (DM), periodically poled lithium niobate (PPLN), fiber-coupled Mach-Zehnder modulator (MZM), fiber amplifier (fAMP), polarizing beam displacer (PBD), and balanced photodetector (BPD). An electro-optic frequency comb with a tooth spacing of 18 GHz is produced on the 1068 nm pump laser through the use of a MZM. This comb then pumps an optical parametric oscillator to produce a  2235 nm optical frequency comb which serves as the coupling laser in these measurements. This comb is combined with the 895 nm probe laser and passed through a cesium vapor cell. The 636 nm dressing laser is counterpropagated through the cell. The probe light passing through the cell is then filtered (to remove the coupling light) before being recorded on a BPD. }
    \label{fig:schematic}
\end{figure}

We utilized an optical parametric oscillator (OPO) to produce the frequency agile, mid-infrared optical frequency comb ~\cite{Long2023Nanosecond, heinigerOPO} which served as the coupling laser (see Fig.~\ref{fig:schematic}).  To generate this optical frequency comb, the 1068 nm OPO seed laser was first passed through a dual-drive Mach-Zehnder modulator \cite{SakamotoEOComb}. By driving both sides of the modulator with an 18 GHz RF tone, an optical frequency comb is produced with a comb spacing of 18 GHz. This comb spacing was selected to roughly match the separation of the selected Rydberg states shown in Fig.~\ref{fig:LevelAndSample}(b). The comb tooth amplitudes were then flattened by adjusting the phase shift and RF power between the two arms of the Mach-Zehnder modulator. This comb was then amplified to 10 W before serving as the pump laser for a commercial, singly resonant, continuous-wave OPO. 

Because only the signal beam is resonant within the OPO cavity, the optical frequency comb on the pump beam was efficiently and coherently transferred on to the mid-infrared idler beam. Critically, this approach allows for the generation of a high power, frequency agile comb, where the comb spacing and span are entirely flexible and can be set to well match an arbitrary set of Rydberg levels. In addition, the output comb can be tuned over a spectral range between 2200 nm and 4000 nm.  For the present measurements it was tuned to 2235 nm, where the comb had a total power $>$2 W and seven strong comb teeth which had an average power per tooth near 300 mW. This power per comb tooth was more than sufficient to allow for multi-state Rydberg preparation.   

The optical frequency comb coupling laser was then combined with the 895 nm probe laser and passed through a $^{133}$Cs filled vapor cell. The 636 nm dressing laser was counter-propagated through the cell in order to reduce the Doppler effect. The transmitted probe beam was then spectrally filtered (in order to remove the coupling light) and measured on a balanced photo-diode, allowing for near shot-noise-limited detection.

\section{Broadband Reception}
\begin{figure}[ht]
    \centering
    \includegraphics[width=\columnwidth]{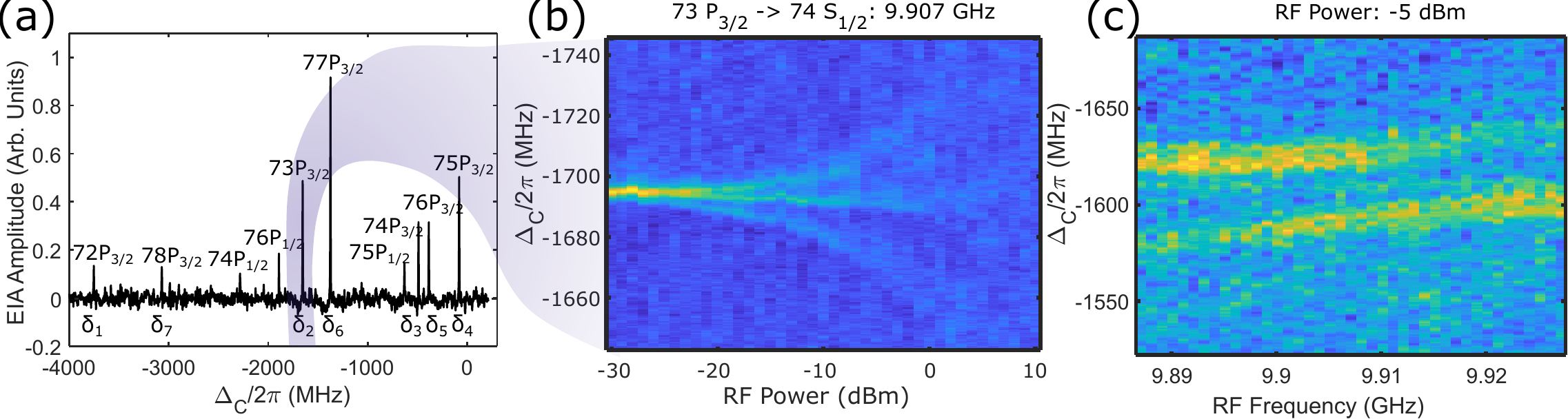}
    \caption{(a) Sample electromagnetically induced absorption (EIA) spectrum showing the peaks corresponding to the different Rydberg states as the coupling laser detuning ($\Delta_C$) is scanned. (b) Response of the 73 P$_{3/2}$ Rydberg state as we increase the RF power of an applied 9.907 GHz field. (c) Corresponding 73 P$_{3/2}$ response as the applied RF frequency is varied near the Rydberg resonance while the applied RF power is fixed to -5 dBm. }
    \label{fig:sample_scan}
\end{figure}

As an initial demonstration we scanned the OPO pump laser (and thus the entire coupling optical frequency comb) over a range of 2 GHz in order to excite each Rydberg state individually, as shown in Fig.~\ref{fig:sample_scan} (a). The detuning of each state corresponds to the energy difference between the nearest comb tooth and the state in question (as defined in Table~\ref{tab:detunings}). The peaks in the spectrum show the seven Rydberg states which can be prepared by the optical frequency comb coupling laser. We note that the reverse configuration can also be performed. By locking the coupling laser and tuning the comb spacing and tooth amplitudes, we can achieve fast switching between the distinct Rydberg states and employ unique demodulation schemes to receive information from the various frequency ranges simultaneously. 

Armed with this multi-state readout, we can utilize this approach to observe a broad range of effects on the different Rydberg states (see Fig.~\ref{fig:LevelAndSample} (c)) being probed. For example, we can demonstrate the independent response of selected states to particular RF fields, the splitting of the Rydberg state resonances for calibration purposes, and even a Stark shift when the applied field is strong enough. Figure~\ref{fig:sample_scan} (b) shows the Autler-Townes splitting as the power of the signal generator is increased. By extracting the splitting, we can determine the field strength of the incident field~\cite{6910267},
\begin{equation}
    E = \frac{2\pi\hbar\Delta_{meas}}{\wp_{i,j}}
\end{equation}
where $\hbar$ is the reduced Planck constant, $E$ is the electric field, $\Delta_{meas}$ is the measured Rydberg state splitting, and $\wp_{i,j}$ is the transition dipole moment between the two Rydberg states. 
For example, at -5 dBm of applied RF, we observe a splitting of 30 MHz. 
The transition dipole moment for this transition is 2480 $e^{-1}a_0^{-1}$, where $e$ is the electron charge and $a_0$ is the Bohr radius. 
For this case, the field measured by the atoms was 0.95 V/m.

\begin{table}[ht]
\centering
\begin{tabular}{|l |l |l |l |l |l |l |l |} \hline   
 & 72P & 73P & 74P & 75P & 76P & 77P & 78P \\ \hline 
Energy/h (GHz) & -57.7 & -37.6 & -18.4 & 0 & 17.6 & 34.6 & 50.9 \\ \hline 
Comb Energy/h (GHz) & -54 & -36 & -18 & 0 & 18 & 36 & 54 \\ \hline 
 & $\delta_1$ & $\delta_2$& $\delta_3$& $\delta_4$& $\delta_5$& $\delta_6$& $\delta_7$\\ \hline 
Detuning from  Tooth (GHz) & -3.7 & -1.6 & -0.4 & 0 & -0.3 & -1.3 & -3.0 \\ \hline
\end{tabular}
\caption{Energy separation of the different Rydberg P$_{3/2}$ states in the manifold. Also shown are the relative comb spacing and detunings of these states from the nearest comb teeth.}
\label{tab:detunings}
\end{table}

Another effect which can be observed are avoided crossings as the frequency of the applied RF is scanned (see Fig.~\ref{fig:sample_scan} (c)). The expected AT behavior with RF detuning is observed, in that there are three main effects on the observed splitting of the electromagnetically induced transparency (EIT) signal~\cite{10.1063/1.4947231}: (1) the two peaks of the EIT signal are nonsymmetric, (2) the separation between the two AT peaks increases with RF detuning, and (3) one peak is pulled to the zero detuning location of the coupling laser. In this case, the splitting is proportional to the generalized Rabi frequency,
\begin{equation}
    \Delta_{meas} = \sqrt{\Omega_{RF}^2+\Delta_{RF}^2},
    \label{eq:gen_rabi}
\end{equation}
where $\Omega_{RF}$ is proportional to the field strength, $\Delta_{meas}$ is the measured splitting for a given RF detuning, and $\Delta_{RF}$ is the detuning of the RF from the atomic resonance. As we perform a broad scan in frequency, we can use these avoided crossings to identify the location of the different RF Rydberg transitions.

\section{Results and Discussion}

The unique broadband nature of the present method becomes apparent when different RF fields are applied to the atoms. For these measurements we used three horn antennas to apply RF fields ranging from 1 GHz to 40 GHz.  The lower RF bound was set by the atomic transitions between states in this region of the Rydberg manifold while the upper bound was set by the available RF sources and antennas. We note that the use of higher frequency RF sources and antennas should allow for a significantly wider measurable RF range. 

\begin{figure}[ht]
    \centering
    \includegraphics[width=\textwidth]{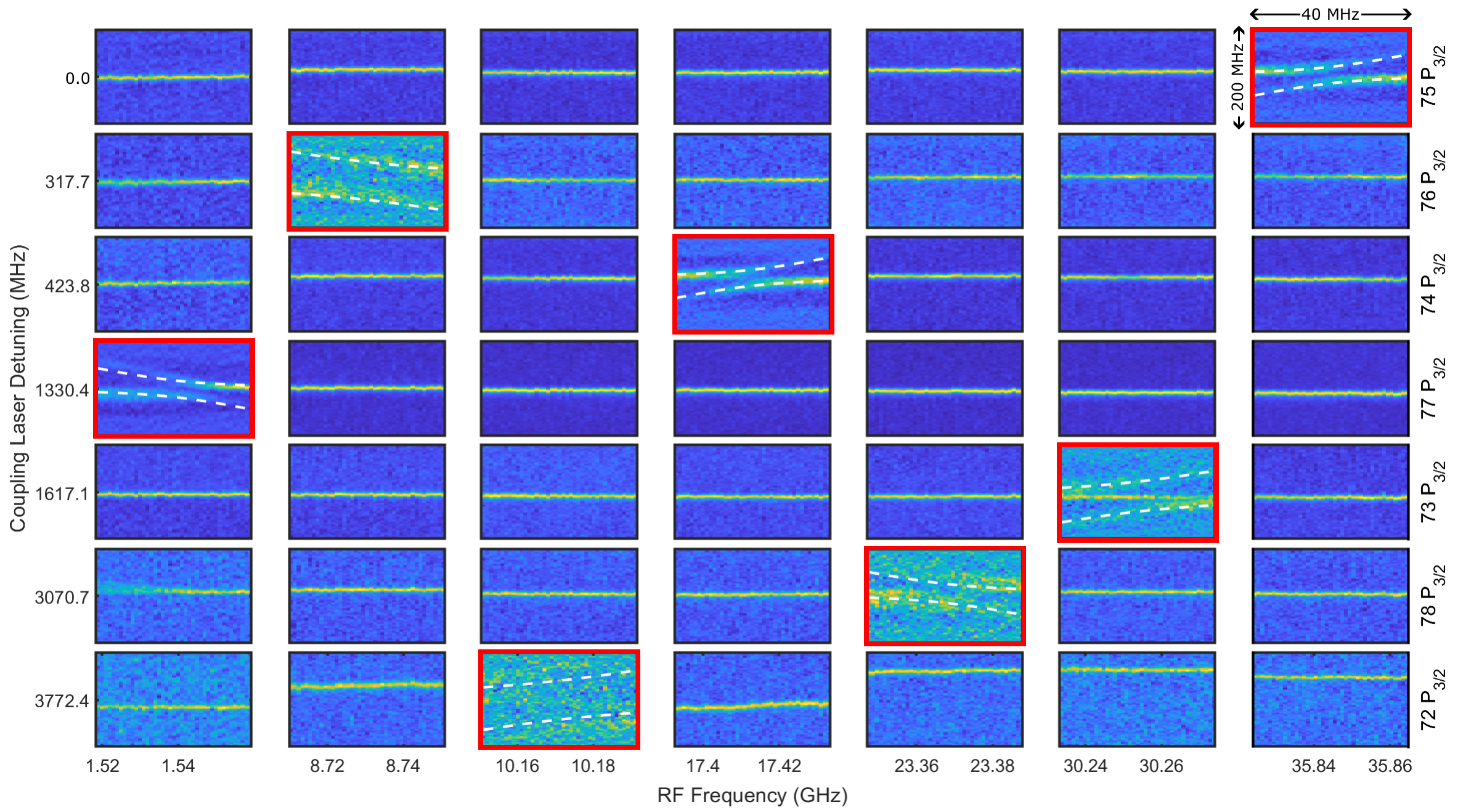}
    \caption{Each panel shows the response of a given Rydberg state to a range of applied radiofrequencies. Each row corresponds to a particular Rydberg state (labeled on the right) and each column corresponds to different applied radiofrequencies. In each panel the y-axis is the coupling laser detuning used to observe the given Rydberg state and the x-axis is a scan of the RF frequency centered on the atomic resonance with a span of 40 MHz. We have highlighted the panels in which we would expect to see a response.}
    \label{fig:freq_scan}
\end{figure}

For these measurements we applied RF frequencies over 20-MHz-wide ranges which were centered on Rydberg transitions between 1 GHz and 40 GHz (see Figure ~\ref{fig:freq_scan}). For example, the first column in Fig.~\ref{fig:freq_scan} shows the response of the seven states to a field ranging from 1.440 GHz to 1.480 GHz. A complete table of each of the different transition frequencies we can access is given in the Supplementary Material Table S1. For each column in Fig.~\ref{fig:freq_scan} we highlight the panel where we expect a response to the RF field. We observed avoided crossings caused by the mixing of the optically excited Rydberg states and the Rydberg state coupled by the RF field ~\cite{Foot2005Atomic}. 

The fits shown in the highlighted panels map the response of the Rydberg state energies using the generalized Rabi in Eq.~\ref{eq:gen_rabi}. The energy of each dressed state present in AT splitting is given by difference and sum of the generalized Rabi frequency and the detuing of the RF frequency from the atomic resonance,
\begin{equation}
    \Delta_{\pm} = \frac{\Delta_{RF}\pm\sqrt{\Omega_{RF}^2+\Delta_{RF}^2}}{2}
\end{equation}
where $\Delta_{\pm}$ is the shift for each dressed state in MHz. From each highlighted panel, we extract the transmission for each given RF frequency and peak fit the data to find the energy of the two dressed states. We fit the dressed state energy separation as a function of RF detuning to obtain the fits seen. These fits can also be used to determine the strength of the RF field present at the atoms.

Importantly, since each of the transition frequencies in this section of the Rydberg manifold are separated by over 50 MHz, a single RF frequency is expected to only interact with a single Rydberg state, thus allowing for complete orthogonality. Experimentally this is what we observe, where the measured response to the RF fields in a multitude of different frequency bands up to 40 GHz exhibits no effects of cross-talk between Rydberg states. This orthogonal nature is critical for communication protocols such as frequency hopping shared spectrum (FHSS) and orthogonal frequency domain multiplexing (OFDM).

FHSS is a protocol that relies on frequency-agile sensors that can quickly switch between frequency bands in order to avoid jamming or perform signal interception. 
However, the bands are generally limited by the bandwidth of the antenna being utilized for reception to at most a few of gigahertz. 
Critically, with the present frequency comb-based approach, it should be possible to switch between vastly different operational bands and therefore significantly improve the performance of this protocol for Rydberg sensing.  In particular, by tuning the comb frequency or coupling laser detuning ($\Delta_C$) we should be able to separate the carrier tones. 

Orthogonal frequency division multiplexing (OFDM) is another protocol which relies upon spreading information over several frequency bands simultaneously. In this way information is simultaneously sent over multiple carriers and each carrier must be separately demodulated. While the FHSS protocol can be achieved by utilizing Stark shifting measurements, OFDM requires that all of the carriers be received simultaneously and typically have the same base-band modulation.  This means that if a single probe was used for readout, all the base-bands would interfere if we were to utilize Stark shifting measurements. This is where the present frequency comb method can separate the RF carriers for simultaneous broadband detection.


Further, the present approach provides a pathway toward a dramatic increase in the response bandwidth of Rydberg atom sensors. By using simultaneous state readout (such as ODFM) in concert with our optical frequency comb-based approach, we should be able to demonstrate a bandwidth that scales with the number of comb teeth~\cite{5727897}. With our current demonstration, it should be possible to demonstrate 100 MHz of bandwidth if we incorporate the nP$_{1/2}$ states. 


\section{Conclusion}
The use of an optical frequency comb as the coupling laser for Rydberg electrometry enables simultaneous and orthogonal measurements over a range of states and therefore radiofrequencies. The increased capabilities made possible by this approach are expected to have an extensive impact on communications and sensing. We note that a classical receiver would require numerous different antennas and substantial down-conversion hardware to receive the same signals that this comb-enabled Rydberg receiver can readily record. Further, for the present demonstration we recorded radiofrequencies from 1 GHz to 40 GHz, however, extending this range to 100 GHz is readily possible, whereby the direct conversion nature of Rydberg atoms is anticipated to be truly enabling as telecommunications continue to push the limits of existing electronic hardware.

\section{Methods}
\subsection*{Laser Stabilization}
The 895 nm probe laser frequency was stabilized using saturated absorption spectroscopy to the 6S$_{1/2},F=4\rightarrow$6P$_{1/2},F=3$ transition. A portion of the 895 nm probe and the 636 nm dressing lasers were utilized to generate a two-photon EIT. The 636 nm dressing laser was tuned and locked to the 6P$_{1/2},F=3\rightarrow$9S$_{1/2},F=4$ transition using the two-photon EIT. The 2235 nm coupling laser was free running and scanned over 4 GHz to access the various Rydberg states prepared by the optical frequency comb.

\subsection*{Laser Parameters}
The size of the probe, dressing, and coupling lasers were set to 750 $\mu$m, 800 $\mu$m, and 820 $\mu$m, respectively. The power of the probe, dressing, and coupling lasers were fixed to 1.5 $\mu$W, 11 mW, and 1 W, respectively. All powers were measured before the cell. The corresponding Rabi rates for the probe, dressing, and coupling are then 700*2$\pi$ kHz, 8*2$\pi$ MHz, and 2.6*2$\pi$ MHz respectively.The powers of the probe and dressing lasers were stabilized using acousto-optic modulators. The power output of the coupling laser was not directly stabilized, but the fiber amplifier for the OPO pump was stabilized internally. 

\subsection*{Comb Parameters}
To produce the mid-infrared optical frequency comb we first generated a near-infrared frequency comb which utilized an external-cavity diode laser (ECDL) as its source. The ECDL output was injected into a dual-drive Mach-Zehnder modulator (MZM). An 18 GHz signal generated by a radiofrequency source was split and each side was amplified to near 0.5 W before being sent to each side of the MZM. One side of the MZM had a 3 dB dB RF attenuator placed on it to achieve a flatter comb \cite{SakamotoEOComb}. The output of the MZM was an optical frequency comb with seven strong teeth which were spaced by 18 GHz. This comb was then amplified to 10 W by a fiber amplifier before being injected into a commercial, singly resonant, continuous wave optical parametric oscillator (OPO). The OPO efficiently and coherently transfer the optical frequency comb to 2235 nm with an output power of more than 2 W ~\cite{Long2023Nanosecond, heinigerOPO}.

\subsection*{Radio Frequency Parameters}
To supply RF radiation to the atoms, we utilized three different horn antennas. An ETS 3115 antenna was used to cover 1 GHz to 18 GHz, we used a Narda 638 standard gain horn to cover 18 GHz to 26.5 GHz, and a Narda 637 standard gain horn to cover 26.5 GHz to 40 GHz~\cite{NISTDisclaimer}. For the frequency scans the power of the RF source for the 1 GHz to 18 GHz was fixed to 5 dBm and the powers of the sources from 18 GHz to 40 GHz were fixed to 10 dBm. The difference was to optimize the visibility of the response between the two ranges.

\subsection*{Data Acquisition}
The data was acquired using a high speed oscilloscope set to collect 2.5 gigasamples per second. We scanned the coupling laser over 4 GHz at a scan rate of 30 Hz. We acquired 5 samples as the RF frequency or power was changed. In post-processing, we overlapped the data by using the 75 P$_{1/2}$ state since we did not scan over any resonant frequencies for that state. 

\backmatter


\bmhead{Acknowledgements}
We would like to thank Paul Hale and team from General Dynamics Missions Systems for lending us a vapor cell with embedded plates that was free of impurities and stray charges, thereby reducing the inhomogeneous broadening at high principle quantum numbers.

\section*{Declarations}


\begin{itemize}
\item This project was partially funded by the NIST on a chip program.

\item Conflict of interest/Competing interests 
A provisional patent has been filed on these results on which N. P., D. A. L., A. B. A.-G., and C. L. H. are co-inventors. 
\item Data availability 
The data underlying the results presented herein will be available at data.nist.gov.
\end{itemize}

\noindent

\bigskip


\begin{appendices}

\section{List of Transition Frequencies}\label{APP:List of Tables}
\begin{figure}[ht]
    \centering
    \includegraphics[width=\textwidth]{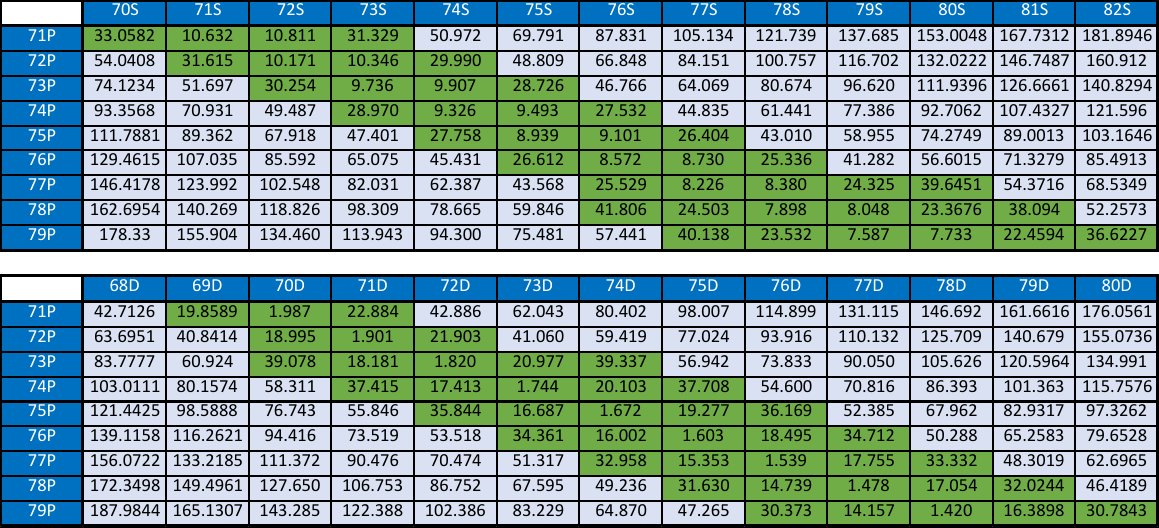}
    \caption{Transition frequencies between the Rydberg states that we excite to and the nearby S and D Rydberg states.}
    \label{fig:enter-label}
\end{figure}





\end{appendices}

\bibliographystyle{sn-nature}
\bibliography{atom_probe_bib}

\begin{thebibliography}{10}
\bibitem{NISTDisclaimer}
\bibinfo{note}{Certain equipment, instruments, software, or materials are identified in this paper in order to specify the experimental procedure adequately. Such identification is not intended to imply recommendation or endorsement of any product or service by NIST, nor is it intended to imply that the materials or equipment identified are necessarily the best available for the purpose.}
\end{thebibliography}

\end{document}